\documentstyle[epsf,axodraw]{article}
\begin{document}

\bigskip
{\Large\bf
\centerline{QCD predictions for annihilation decays of P-wave}
\centerline{quarkonia to next-to-leading order in $\alpha_s$} 
\bigskip
\normalsize

\centerline{Han-Wen Huang$^{1,2}$~~~~Kuang-Ta Chao$^{1,3}$}
\centerline{\sl $^1$ CCAST (World Laboratory), Beijing 100080, P.R.China} 
\centerline{\sl $^2$ Institute of Theoretical Physics, Academia Sinica,
      P.O.Box 2735,}
\centerline{\sl Beijing 100080, P.R.China}
\centerline{\sl $^3$ Department of Physics, Peking University, Beijing 100871,
           P.R.China}
\bigskip

\begin{abstract}
The decay rates of P-wave heavy quarkonia to light hadrons are presented to
leading order in $v^2$ and next-to-leading order in $\alpha_s$. They include
contributions from both the color-singlet component and the color-octet 
component of quarkonia. Applying these results to charmonium and using 
measured decay rates for the $\chi_{c1}$ and $\chi_{c2}$ by E760, we 
determine the two nonperturbative decay matrix elements, and then predict 
the hadronic decay rates of $\chi_{c0}$ and $h_c$, and the electromagnetic 
decay rates of $\chi_{c0}$ and $\chi_{c2}$. The obtained decay rates of 
$\chi_{c0}\rightarrow LH$ and $\chi_{c0}\rightarrow\gamma\gamma$ are in   
agreement with the Crystal Ball result, and also with the new measurement by 
BES. However, the results for $\Gamma(\chi_{c0}\rightarrow LH)$ are dependent 
on the choice of renormalization scale.
\end{abstract}

\vfill\eject\pagestyle{plain}\setcounter{page}{1}

The study of heavy quarkonium physics can provide very interesting tests of 
perturbative quantum chromodynamics (PQCD). Calculations of the rates for 
heavy quarkonium decay into light hadrons were among the early applications of 
PQCD. These early calculations are based on a naive factorization assumption 
that all long-distance nonperturbative effects can be factored into the 
nonrelativistic wavefunction of color singlet $Q\bar{Q}$ or its derivative at 
the origin, and the perturbative part is related to the annihilation rates of 
color-singlet $Q\bar{Q}$ which can be calculated using PQCD. In the 
nonrelativistic limit, this early factorization formalism was supported by 
explicit calculations for S-wave decays at next-to-leading order in 
$\alpha_s$ \cite{hag}. But in the case of P-wave \cite{Barbi} quarkonium decays,
infrared divergences appeared in the perturbative calculations of 
color-singlet $Q\bar{Q}$ annihilation amplitudes. These are clear 
indications that the decay rates are sensitive to nonperturbative effects 
beyond those related to the wavefunction of color-singlet $Q\bar{Q}$ pair or
its derivative at the origin, and not all nonperturbative effects can be 
factored into the color-singlet component of quarkonium. Recently, Bodwin, 
Braaten and Lepage (BBL) have developed a rigorous factorization formalism 
\cite{BBL} which is based on an effective field theory, nonrelativistic 
QCD (NRQCD). This factorization formalism provides a clean separation between 
short-distance effects and long-distance effects for the decay rates and 
production cross sections of heavy quarkonium. 

Nowadays there is a renewed interest in studying the decay of P-wave
charmonium, not only due to the theoretical development mentioned above
but also due to recent experimental results such as the total decay widths
of $\chi_{cJ}$ and the observation of $h_c$. BBL have applied the new 
factorization approach in a phenomenological analysis of P-wave charmonium
decays \cite{pwave}. They give a leading order result with both the 
color-singlet and color-octet $Q\bar{Q}$ components. The next-to-leading 
order correction to the decay of $h_c$ is given in \cite{huang}, where both 
the color-singlet and color-octet contributions are included and the explicit 
cancellation of previously encountered infrared divergence is revealed. 
In \cite{mang} the next-to-leading order color-singlet terms are considered in 
a phenomenological analysis of hadronic annihilation decays of $\chi_{cJ}$. 
Recently the next-to-leading order color-octet corrections to hadronic $\chi_J$
decays have also been calculated \cite{3s1}. In this paper we will perform
a phenomenological study for the hadronic decays of four P-wave charmonium
states by using the results that completely include the next-to-leading order
QCD corrections.

We start with the formulas for the P-wave quarkonium decay widths in the new 
factorization formalism
\begin{equation}
\label{3pj}
\Gamma(\chi_J\rightarrow LH)=2Imf_1(^3P_J)H_1
+2Imf_8(^3S_1)H_8+O(v^2\Gamma),
\end{equation}
\begin{equation}
\label{1p1}
\Gamma(h\rightarrow LH)=2Imf_1(^1P_1)H_1
+2Imf_8(^1S_0)H_8+O(v^2\Gamma),
\end{equation}
where $H_1$ and $H_8$ are the matrix elements of color-singlet and color-octet
operators respectively. The short-distance coefficients can be extracted by 
matching the imaginary part of the on-shell $Q\bar{Q}$ pair forward scattering 
amplitude calculated in full perturbative QCD with that calculated in NRQCD.
We list the results to next-to-leading order in $\alpha_s$ as the following
\begin{eqnarray}\label{cp0}\nonumber
Imf_1(^3P_0)&=&(Imf_1(^3P_0))_0\{1+\frac{\alpha_s}{\pi}[(4b_0-\frac{4n_f}{27}
)ln\frac{\mu}{2m}\\&+&(\frac{454}{81}-\frac{\pi^2}{144})C_A+(-\frac{7}{3}
+\frac{\pi^2}{4})C_F-\frac{58}{81}n_f]\},
\end{eqnarray}
\begin{equation}\label{cp1}
Imf_1(^3P_1)=(Imf_1(^3P_0))_0\frac{\alpha_s}{\pi}[-\frac{4n_f}{27}
ln\frac{\mu}{2m}+(\frac{587}{54}-\frac{317\pi^2}{288})-\frac{16n_f}{81}],
\end{equation}
\begin{eqnarray}\label{cp2}\nonumber
Imf_1(^3P_2)&=&(Imf_1(^3P_2))_0\{1+\frac{\alpha_s}{\pi}[(4b_0-\frac{5n_f}{9}
)ln\frac{\mu}{2m}\\&+&(\frac{2239}{216}-\frac{337\pi^2}{384}+\frac{5ln2}{3})C_A
-4C_F-\frac{29}{27}n_f]\},
\end{eqnarray}
\begin{equation}
\label{IMP}
Imf_1(^1P_1)=\frac{(N_c^2-4)C_F\alpha_s^3}{3N_c^2}(\frac{7\pi^2-118}{48}
-ln\frac{\mu}{2m}),
\end{equation}
\begin{eqnarray}\label{IMS1}\nonumber
Imf_8(^3S_1)&=&(Imf_8(^3S_1))_0\{1+\frac{\alpha_s}{\pi}[4b_0ln\frac{\mu}{2m}
-\frac{5}{9}n_f\\&+&(\frac{133}{18}+\frac{2}{3}ln2-\frac{\pi^2}{4})C_A
-\frac{13}{4}C_F+\frac{5}{n_f}(-\frac{73}{4}+\frac{67}{36}\pi^2)]\},
\end{eqnarray}
\begin{equation}
\label{IMS}
Imf_8(^1S_0)=(Imf_8(^1S_0))_0\{1+\frac{\alpha_s}{\pi}[4b_0ln\frac{\mu}{2m}
-\frac{8}{9}n_f+(\frac{\pi^2}{4}-5)C_F+(\frac{479}{36}-\frac{17\pi^2}{24})C_A]\}
\end{equation},
where
$$
b_0=\frac{1}{12}(11C_A-2n_f),
$$
and $C_F=\frac{N_c^2-1}{2N_c}$,~~$C_A=N_c$. The coefficients $Imf_1(^3P_1)$ starts 
in order $\alpha_s^3$, hence they only are given to leading order; while all other 
coefficients start in order $\alpha_s^2$, whose next-to-leading order corrections
are also given. We note that (\ref{IMP}) and (\ref{IMS}) are
given in \cite{huang} and (\ref{IMS1}) is given in \cite{3s1}. Coefficients $Imf_1(^3P_J)$ in (\ref{cp0})-(\ref{cp2}) have been calculated in \cite{Barbi}
and listed in \cite{3s1}, where quark and antiquark are taken off-shell and binding energy regularization scheme was used. Here we recalculate them 
by using dimensional regularization to control the infrared divergence, and
there are some differences between our results (\ref{cp0})-(\ref{cp2})
and that in \cite{3s1}. This difference only comes from the diagram in Fig.1 which 
represents the inclusive processes $Q\bar{Q}(^3P_J)\rightarrow q_i\bar{q}_ig$.
In the $d=4-2\epsilon$ dimension space, contributions of the three particle cut
diagram in Fig.1 to the imaginary part of $Q\bar{Q}(^3P_J)$ pair scattering 
amplitude are
\begin{equation}\label{3p0}
Im{\cal M}^{full~QCD}_{Fig.1}(Q\bar{Q}(^3P_0))=(Imf_8(^3S_1))_0f(\epsilon)
\frac{4C_F\alpha_s}{3N_C\pi}(-\frac{1}{2\epsilon_{IR}})+(Imf_1(^3P_0))_0
\frac{\alpha_s}{\pi}(-\frac{58n_f}{81}-\sum_i\frac{2}{3}ln\frac{m_i}{2m}),
\end{equation}
\begin{equation}\label{3p1}
Im{\cal M}^{full~QCD}_{Fig.1}(Q\bar{Q}(^3P_1))=(Imf_8(^3S_1))_0f(\epsilon)
\frac{4C_F\alpha_s}{3N_C\pi}(-\frac{1}{2\epsilon_{IR}})+(Imf_1(^3P_0))_0
\frac{\alpha_s}{\pi}(-\frac{16n_f}{81}),
\end{equation}
\begin{equation}\label{3p2}
Im{\cal M}^{full~QCD}_{Fig.1}(Q\bar{Q}(^3P_2))=(Imf_8(^3S_1))_0f(\epsilon)
\frac{4C_F\alpha_s}{3N_C\pi}(-\frac{1}{2\epsilon_{IR}})+(Imf_1(^3P_2))_0
\frac{\alpha_s}{\pi}(-\frac{29n_f}{27}-\sum_i\frac{2}{3}ln\frac{m_i}{2m}).
\end{equation}
Here
$$
f(\epsilon)=(\frac{4\pi\mu^2}{4m^2})^{\epsilon}\Gamma(1+\epsilon)
$$
The results coming from the diagrams that represent the inclusive processes
$Q\bar{Q}(^3P_J)\rightarrow gg$ and $Q\bar{Q}(^3P_J)\rightarrow ggg$ are finite
and have been given in \cite{Barbi}.

\begin{center}\begin{picture}(120,100)(0,0)
\Line(20,80)(40,80)\Line(40,80)(40,20)\Line(40,20)(20,20)
\Photon(40,80)(50,80){1}{4}\CArc(60,80)(10,0,360)
\Photon(70,80)(80,80){1}{4}\Photon(40,20)(80,20){1}{8}
\Line(80,80)(100,80)\Line(80,20)(80,80)\Line(100,20)(80,20)
\end{picture}

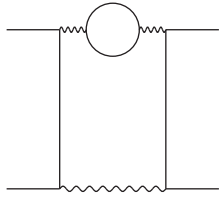
{\sl\bf Fig.1~~Feynman diagram with three particle cut contributing to the 
divergence terms in the full theory calculation of $Q\bar{Q}$ annihilation
amplitudes}
\end{center}

While in the effective field theory NRQCD, the corresponding scattering 
amplitudes can be written as
\begin{equation}\label{3pn}
Im{\cal M}(^3P_J)_{NRQCD}=\frac{Imf_1(^3P_J)}{m^6}+(Imf_8(^3S_1))_0
\frac{4C_F\alpha_s}{3m^6N_c\pi}(-\frac{1}{2\epsilon_{IR}}
+\frac{1}{2\epsilon_{UV}}).
\end{equation} 
Comparing (\ref{3pn}) with (\ref{3p0})--(\ref{3p2}), it is obvious that the 
divergence terms are removed and finite coefficients $Imf_1(^3P_J)$ 
(\ref{cp0})---(\ref{cp2}) can be obtained. It is important to point out that 
if one replaces $ln\frac{m}{\varepsilon}$ in the expressions in \cite{3s1} by 
$-\frac{1}{2\epsilon_{IR}}$, then the divergent terms are the same 
as those in (\ref{3p0})---(\ref{3p2}).
The difference only occurs in their finite terms due to different regularization
scheme being used. It is certainly true that the coefficients of 4-fermion 
operators
must be infrared finite and independent of the choice of regularization 
procedures because all nonperturbative effects are factored into the matrix 
elements. Note that the coefficients can be derived consistently only by 
taking the same regularization scheme in full QCD and in effective NRQCD. The 
advantage of using dimensional regularization is that the on-shell condition 
and gauge invariance are maintained manifestly and conventional treatment of 
NRQCD is under the on-shell condition, thus we
can give an explicit cancellation for divergences appeared previously. The
introduction of off-shell binding energy makes it difficult to do calculation
in NRQCD, and the results are incomplete if simply absorbing the divergences
associated with the logarithm of binding energy into the matrix elements
of color-octet operators.  

Now we apply the factorization formula to charmonium systems. For the lowest
radial excitation, the $^3P_J$ states are called $\chi_{cJ}$ and the $^1P_1$
state is called $h_c$. The explicit form for their decay rates into
light hadrons at leading order in $v^2$ are
\begin{equation}\label{c0}
\Gamma(\chi_{c0}\rightarrow LH)=C_{00}\alpha_s^2(m_c)(1+C_{01}\frac{\alpha_s}
{\pi})H_1+D_{0}\alpha_s^2(m_c)(1+D_{1}\frac{\alpha_s}{\pi})H_8(m_c),
\end{equation}
\begin{equation}\label{c1}
\Gamma(\chi_{c1}\rightarrow LH)=C_1\alpha_s^3H_1
+D_0\alpha_s^2(m_c)(1+D_1\frac{\alpha_s}{\pi})H_8(m_c),
\end{equation}
\begin{equation}\label{c2}
\Gamma(\chi_{c2}\rightarrow LH)=C_{20}\alpha_s^2(m_c)(1+C_{21}\frac{\alpha_s}
{\pi})H_1+D_0\alpha_s^2(m_c)(1+D_1\frac{\alpha_s}{\pi})H_8(m_c),
\end{equation}
\begin{equation}\label{hc}
\Gamma(h_c\rightarrow LH)=C^{\prime}_1\alpha_s^3H_1
+D^{\prime}_0\alpha_s^2(m_c)(1+D^{\prime}_1\frac{\alpha_s}{\pi})H_8(m_c),
\end{equation}
where ``LH" on the left hand of (\ref{c0})---(\ref{hc}) represents all final states
consisting of light hadrons, and the coefficients are 
$$
C_{00}=\frac{4\pi}{3},~~C_{01}=8.710;
$$
$$
C_1=-0.370,~~C^{\prime}_1=-0.161;
$$
$$
C_{20}=\frac{16\pi}{45},~~C_{21}=-5.061;
$$
$$
D_0=\pi,~~D_1=4.110;
$$
$$
D^{\prime}_0=\frac{5\pi}{6},~~D^{\prime}_1=6.66.
$$
In deriving these coefficients we have taken $N_c=3,~n_f=3$ and made a choice 
$\mu=m_c$ for the scale in the $\overline{MS}$ scheme. The large size of some 
coefficients for the correction terms is apparent. These numbers obviously depend 
on the definition of the renormalized couplings $\alpha_s$. We will study the 
the renormalization scale dependence of the results later. In the following we use measured 
decay rates of the $\chi_{c1}$ and $\chi_{c2}$ to predict the inclusive decay 
rates of the $\chi_{c0}$ and $h_c$, and the theoretical uncertainties will 
be estimated by considering relativistic corrections and high order 
perturbative QCD corrections.

Precision measurements of the total decay rates of the $^3P_1$ state 
$\chi_{c1}$ and $^3P_2$ state $\chi_{c2}$ have recently been carried out
at Fermilab by the E760 collaboration. Their results with statistical and
systematic errors are \cite{particle}
$$
\Gamma(\chi_{c2})=2.00\pm 0.18Mev,
$$
$$
\Gamma(\chi_{c1})=0.88\pm 0.14Mev.
$$
It is well known that the main decay modes of these P-wave charmonium states 
are the decay into light hadrons and the radiative transitions into $J/\psi$ 
or $\eta_c$.
Other decay modes such as pionic transitions of the P states to the S states,
of which the most important decay modes should be $J/\psi+\pi\pi$ and
$\eta_c+\pi\pi$, contribute much less to the total widths and therefore can be 
neglected \cite{kuang}.
Previous experiments have measured the branching fractions for the radiative
transitions of the $\chi_{c1}$ and $\chi_{c2}$ into the $J/\psi$, and they are
$B(\chi_{c1}\rightarrow\gamma J/\psi)=0.273\pm0.016$, and $B(\chi_{c2}\rightarrow\gamma J/\psi)=0.135\pm0.011$ \cite{particle}.
We use the radiative branching fractions and the total decay rates to obtain
the partial rates for light hadronic decays of the $\chi_{c1}$ and $\chi_{c2}$
\begin{equation}\label{dc1}
\Gamma(\chi_{c1}\rightarrow LH)=0.64\pm 0.10Mev,
\end{equation}
\begin{equation}\label{dc2}
\Gamma(\chi_{c2}\rightarrow LH)=1.71\pm 0.16Mev.
\end{equation}

$H_1$ and $H_8$ can be obtained directly by using (\ref{c1})
and (\ref{c2}),
\begin{equation}\label{h1}
H_1=\frac{\Gamma(\chi_{c2}\rightarrow LH)-\Gamma(\chi_{c1}\rightarrow LH)}
{C_{20}\alpha_s^2(1+C_{21}\frac{\alpha_s}{\pi})-C_1\alpha_s^3},
\end{equation}
\begin{equation}\label{h8}
H_8=\frac{\Gamma(\chi_{c2}\rightarrow LH)-C_{20}\alpha_s^2(1+C_{21}\frac{\alpha_s}{\pi})H_1}
{D_0\alpha_s^2(1+D_1\frac{\alpha_s}{\pi})}.
\end{equation}
Here we determine $\alpha_s(m_c)$ by taking the coupling constant 
$\alpha_s(m_b)=0.189\pm 0.008$ extracted from bottonium decays and evolving
it down to the scale $m_c$. The resulting value of the coupling constant is
$\alpha_s(m_c)=0.29\pm 0.02$. Inserting (\ref{dc1}) and (\ref{dc2}) into (\ref{h1})
and (\ref{h8}), we obtain
$$
H_1=18.4\pm 5.2Mev,
$$
$$
H_8=2.21\pm 0.15Mev.
$$
The ratio of the two nonperturbative parameters is $H_8/H_1\approx 0.10$,
while it was determined to be $0.21$ \cite{pwave} if it was considered only 
to leading order in $\alpha_s$. 
Substituting $H_1$ and $H_8$ into (\ref{c0}) and (\ref{hc}) we can 
easily get the decay widths of $\chi_{c0}$ and $h_c$ into light hadrons
\begin{equation}
\Gamma(\chi_{c0}\rightarrow LH)=12.4\pm 3.2Mev,
\end{equation}
\begin{equation}
\Gamma(h_c\rightarrow LH)=0.71\pm 0.07Mev.
\end{equation}
Adding the radiative decay rate for $\chi_{c0}$ whose branching fraction
has been measured to be $(0.66\pm 0.18)\%$, we obtain the total decay rate
$\Gamma(\chi_{c0})=12.5\pm 3.2Mev$, which agrees with the earlier Crystal
Ball value $14\pm 5Mev$ \cite{particle}, also with the new (preliminary)
value of $\Gamma_{tot}(\chi_{c0})=15.0^{+3.2}_{-2.8}Mev$ measured by BES,
using $3.5\times 10^6~\psi^{\prime}(\psi(3686))$ events in $\psi^{\prime}
\rightarrow\gamma\chi_{c0}$ and $\chi_{c0}\rightarrow\pi^+\pi^-,K^+K^-$
decay channels \cite{gu}.
The rate for decay $h_c\rightarrow\gamma\eta_c$ has been 
estimated within a phenomenological framework to be about $(340-380)kev$ 
\cite{chao}, 
and the total decay widths of $h_c$ is then $\Gamma(h_c)\approx 1.07Mev$, which is also consistent with the experimental result \cite{1p1}.

For the electromagnetic decays to next to leading order in $\alpha_s$, we have
\begin{equation}
\Gamma(\chi_{c0}\rightarrow\gamma\gamma)=6\pi e^4_c\alpha^2[1+(\frac{\pi^2}{3}
-\frac{28}{9})\frac{\alpha_s}{\pi}]H_1,
\end{equation}
\begin{equation}
\Gamma(\chi_{c2}\rightarrow\gamma\gamma)=\frac{8\pi}{5}e^4_c\alpha^2(1
-\frac{16\alpha_s}{3\pi})H_1.
\end{equation}
With the determined value for $H_1$, and $e_c=2/3,~\alpha=1/137,~\alpha_s
=0.29$, we predict
\begin{equation}
\Gamma(\chi_{c0}\rightarrow\gamma\gamma)=(3.72\pm 1.11)kev,
\end{equation}
which is in agreement with the observed value $(4.0\pm 2.8)kev$ by Crystal
Ball \cite{particle}, and also predict
\begin{equation}
\Gamma(\chi_{c2}\rightarrow\gamma\gamma)=(0.49\pm 0.15)kev,
\end{equation}
which is larger than the E760 value \cite{E760} but smaller than the CLEO
value \cite{CLEO}.

It is important to have a reasonable estimate for the theoretical uncertainties in our 
results. The two main sources of theoretical errors are relativistic corrections and 
higher-order perturbative corrections. Our formula (\ref{c0})--(\ref{hc}) are only valid 
to leading order in $v^2$ and high order relativistic corrections are not known at 
present. The error due to neglecting high order relativistic contributions could be of order 
$v^2\approx 30\%$. On the other hand, we find that the one-loop coefficients in 
(\ref{c0})--(\ref{hc}) are very large and strongly depend on the scale $\mu$. It is well 
known that when working to all order in $\alpha_s$, the decay rates, being the physical 
observables, will not rely on the choice of $\mu$. However we only do calculation to 
next-to-leading order in $\alpha_s$, therefore the analyses for the scale dependence of 
the results and the estimates for the higher-order effects are needed. The results are 
shown in Fig.2 and Fig.3 for decay rates $\Gamma(\chi_{c0}\rightarrow LH)$ and 
$\Gamma(h_c\rightarrow LH)$ respectively. For the running coupling constant $\alpha_s$ with two-loops, three 
values $\Lambda^{(3)}_{\overline{MS}}=200Gev,~250Gev,~and 300Gev$ are used. The pictures show 
that our results are quite stable in the case of large $\mu$, say, $\mu>2m_c$. In the physically
motivated range 
$\mu=m_c$ to $2m_c$, the decay rates vary from 15Mev to 9Mev for 
$\Gamma(\chi_{c0}\rightarrow LH)$ and from 0.7Mev to 0.6Mev for $\Gamma(h_c\rightarrow LH)$
respectively, while the obtained two phenomenological parameters $H_1=22.0-19.0Mev,~H_8=2.3-3.1Mev$
and the ratio $H_8/H_1=0.11-0.16$. 
It is interesting to note that, although there are large theoretical
uncertainties due to the scale dependence and higher-order QCD corrections, our estimate for
$\Gamma(\chi_{c0}\rightarrow LH)$ is enhanced greatly compared with the previous leading
order result $\Gamma_{tot}(\chi_{c0})=(4.8\pm 0.7)Mev$ \cite{pwave}, which is smaller by 
a factor of three than the experimental value.  

\begin{figure}
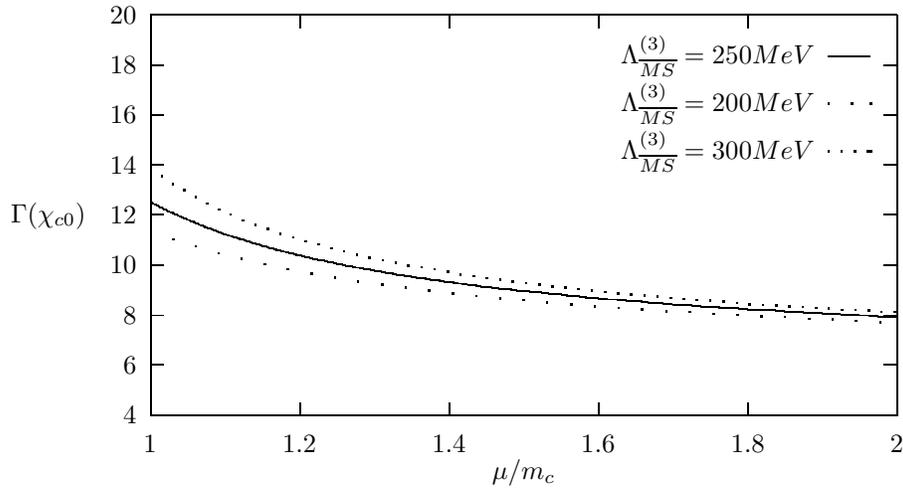

\input fig1.tex
\caption{Renormalization scale dependence of the decay width 
$\Gamma(\chi_{c0}\rightarrow LH)$} 
\end{figure}

\begin{figure}
\input fig2.tex
\caption{Renormalization scale dependence of the decay width $\Gamma(h_c\rightarrow LH)$}
\end{figure}
  
\begin{figure}
\setlength{\unitlength}{0.240900pt}
\ifx\plotpoint\undefined\newsavebox{\plotpoint}\fi
\sbox{\plotpoint}{\rule[-0.175pt]{0.350pt}{0.350pt}}%
\begin{picture}(1500,900)(0,0)
\tenrm
\sbox{\plotpoint}{\rule[-0.175pt]{0.350pt}{0.350pt}}%
\put(264,158){\rule[-0.175pt]{4.818pt}{0.350pt}}
\put(242,158){\makebox(0,0)[r]{10}}
\put(1416,158){\rule[-0.175pt]{4.818pt}{0.350pt}}
\put(264,315){\rule[-0.175pt]{4.818pt}{0.350pt}}
\put(242,315){\makebox(0,0)[r]{15}}
\put(1416,315){\rule[-0.175pt]{4.818pt}{0.350pt}}
\put(264,473){\rule[-0.175pt]{4.818pt}{0.350pt}}
\put(242,473){\makebox(0,0)[r]{20}}
\put(1416,473){\rule[-0.175pt]{4.818pt}{0.350pt}}
\put(264,630){\rule[-0.175pt]{4.818pt}{0.350pt}}
\put(242,630){\makebox(0,0)[r]{25}}
\put(1416,630){\rule[-0.175pt]{4.818pt}{0.350pt}}
\put(264,787){\rule[-0.175pt]{4.818pt}{0.350pt}}
\put(242,787){\makebox(0,0)[r]{30}}
\put(1416,787){\rule[-0.175pt]{4.818pt}{0.350pt}}
\put(264,158){\rule[-0.175pt]{0.350pt}{4.818pt}}
\put(264,113){\makebox(0,0){1}}
\put(264,767){\rule[-0.175pt]{0.350pt}{4.818pt}}
\put(498,158){\rule[-0.175pt]{0.350pt}{4.818pt}}
\put(498,113){\makebox(0,0){1.2}}
\put(498,767){\rule[-0.175pt]{0.350pt}{4.818pt}}
\put(733,158){\rule[-0.175pt]{0.350pt}{4.818pt}}
\put(733,113){\makebox(0,0){1.4}}
\put(733,767){\rule[-0.175pt]{0.350pt}{4.818pt}}
\put(967,158){\rule[-0.175pt]{0.350pt}{4.818pt}}
\put(967,113){\makebox(0,0){1.6}}
\put(967,767){\rule[-0.175pt]{0.350pt}{4.818pt}}
\put(1202,158){\rule[-0.175pt]{0.350pt}{4.818pt}}
\put(1202,113){\makebox(0,0){1.8}}
\put(1202,767){\rule[-0.175pt]{0.350pt}{4.818pt}}
\put(1436,158){\rule[-0.175pt]{0.350pt}{4.818pt}}
\put(1436,113){\makebox(0,0){2}}
\put(1436,767){\rule[-0.175pt]{0.350pt}{4.818pt}}
\put(264,158){\rule[-0.175pt]{282.335pt}{0.350pt}}
\put(1436,158){\rule[-0.175pt]{0.350pt}{151.526pt}}
\put(264,787){\rule[-0.175pt]{282.335pt}{0.350pt}}
\put(45,472){\makebox(0,0)[l]{\shortstack{$H_1$}}}
\put(850,68){\makebox(0,0){$\mu/m_c$}}
\put(264,158){\rule[-0.175pt]{0.350pt}{151.526pt}}
\put(1306,722){\makebox(0,0)[r]{$\Lambda^{(3)}_{\overline{MS}}=250MeV$}}
\put(1328,722){\rule[-0.175pt]{15.899pt}{0.350pt}}
\put(264,424){\usebox{\plotpoint}}
\put(264,424){\rule[-0.175pt]{0.873pt}{0.350pt}}
\put(267,423){\rule[-0.175pt]{0.873pt}{0.350pt}}
\put(271,422){\rule[-0.175pt]{0.873pt}{0.350pt}}
\put(274,421){\rule[-0.175pt]{0.873pt}{0.350pt}}
\put(278,420){\rule[-0.175pt]{0.873pt}{0.350pt}}
\put(282,419){\rule[-0.175pt]{0.873pt}{0.350pt}}
\put(285,418){\rule[-0.175pt]{0.873pt}{0.350pt}}
\put(289,417){\rule[-0.175pt]{0.873pt}{0.350pt}}
\put(293,416){\rule[-0.175pt]{1.032pt}{0.350pt}}
\put(297,415){\rule[-0.175pt]{1.032pt}{0.350pt}}
\put(301,414){\rule[-0.175pt]{1.032pt}{0.350pt}}
\put(305,413){\rule[-0.175pt]{1.032pt}{0.350pt}}
\put(310,412){\rule[-0.175pt]{1.032pt}{0.350pt}}
\put(314,411){\rule[-0.175pt]{1.032pt}{0.350pt}}
\put(318,410){\rule[-0.175pt]{1.032pt}{0.350pt}}
\put(322,409){\rule[-0.175pt]{1.164pt}{0.350pt}}
\put(327,408){\rule[-0.175pt]{1.164pt}{0.350pt}}
\put(332,407){\rule[-0.175pt]{1.164pt}{0.350pt}}
\put(337,406){\rule[-0.175pt]{1.164pt}{0.350pt}}
\put(342,405){\rule[-0.175pt]{1.164pt}{0.350pt}}
\put(347,404){\rule[-0.175pt]{1.164pt}{0.350pt}}
\put(352,403){\rule[-0.175pt]{1.397pt}{0.350pt}}
\put(357,402){\rule[-0.175pt]{1.397pt}{0.350pt}}
\put(363,401){\rule[-0.175pt]{1.397pt}{0.350pt}}
\put(369,400){\rule[-0.175pt]{1.397pt}{0.350pt}}
\put(375,399){\rule[-0.175pt]{1.397pt}{0.350pt}}
\put(380,398){\rule[-0.175pt]{1.807pt}{0.350pt}}
\put(388,397){\rule[-0.175pt]{1.807pt}{0.350pt}}
\put(396,396){\rule[-0.175pt]{1.807pt}{0.350pt}}
\put(403,395){\rule[-0.175pt]{1.807pt}{0.350pt}}
\put(411,394){\rule[-0.175pt]{2.329pt}{0.350pt}}
\put(420,393){\rule[-0.175pt]{2.329pt}{0.350pt}}
\put(430,392){\rule[-0.175pt]{2.329pt}{0.350pt}}
\put(439,391){\rule[-0.175pt]{2.329pt}{0.350pt}}
\put(449,390){\rule[-0.175pt]{2.329pt}{0.350pt}}
\put(459,389){\rule[-0.175pt]{2.329pt}{0.350pt}}
\put(468,388){\rule[-0.175pt]{3.493pt}{0.350pt}}
\put(483,387){\rule[-0.175pt]{3.493pt}{0.350pt}}
\put(498,386){\rule[-0.175pt]{3.613pt}{0.350pt}}
\put(513,385){\rule[-0.175pt]{3.613pt}{0.350pt}}
\put(528,384){\rule[-0.175pt]{6.986pt}{0.350pt}}
\put(557,383){\rule[-0.175pt]{6.986pt}{0.350pt}}
\put(586,382){\rule[-0.175pt]{7.227pt}{0.350pt}}
\put(616,381){\rule[-0.175pt]{6.986pt}{0.350pt}}
\put(645,380){\rule[-0.175pt]{35.171pt}{0.350pt}}
\put(791,381){\rule[-0.175pt]{14.213pt}{0.350pt}}
\put(850,382){\rule[-0.175pt]{6.986pt}{0.350pt}}
\put(879,383){\rule[-0.175pt]{14.213pt}{0.350pt}}
\put(938,384){\rule[-0.175pt]{6.986pt}{0.350pt}}
\put(967,385){\rule[-0.175pt]{7.227pt}{0.350pt}}
\put(997,386){\rule[-0.175pt]{6.986pt}{0.350pt}}
\put(1026,387){\rule[-0.175pt]{6.986pt}{0.350pt}}
\put(1055,388){\rule[-0.175pt]{3.493pt}{0.350pt}}
\put(1069,389){\rule[-0.175pt]{3.493pt}{0.350pt}}
\put(1084,390){\rule[-0.175pt]{7.227pt}{0.350pt}}
\put(1114,391){\rule[-0.175pt]{6.986pt}{0.350pt}}
\put(1143,392){\rule[-0.175pt]{6.986pt}{0.350pt}}
\put(1172,393){\rule[-0.175pt]{3.613pt}{0.350pt}}
\put(1187,394){\rule[-0.175pt]{3.613pt}{0.350pt}}
\put(1202,395){\rule[-0.175pt]{6.986pt}{0.350pt}}
\put(1231,396){\rule[-0.175pt]{6.986pt}{0.350pt}}
\put(1260,397){\rule[-0.175pt]{3.613pt}{0.350pt}}
\put(1275,398){\rule[-0.175pt]{3.613pt}{0.350pt}}
\put(1290,399){\rule[-0.175pt]{6.986pt}{0.350pt}}
\put(1319,400){\rule[-0.175pt]{3.493pt}{0.350pt}}
\put(1333,401){\rule[-0.175pt]{3.493pt}{0.350pt}}
\put(1348,402){\rule[-0.175pt]{6.986pt}{0.350pt}}
\put(1377,403){\rule[-0.175pt]{3.613pt}{0.350pt}}
\put(1392,404){\rule[-0.175pt]{3.613pt}{0.350pt}}
\put(1407,405){\rule[-0.175pt]{6.986pt}{0.350pt}}
\sbox{\plotpoint}{\rule[-0.500pt]{1.000pt}{1.000pt}}%
\put(1306,652){\makebox(0,0)[r]{$\Lambda^{(3)}_{\overline{MS}}=200MeV$}}
\put(1350,652){\rule{.1pt}{.1pt}}
\put(264,518){\rule{.1pt}{.1pt}}
\put(293,511){\rule{.1pt}{.1pt}}
\put(323,505){\rule{.1pt}{.1pt}}
\put(352,500){\rule{.1pt}{.1pt}}
\put(381,495){\rule{.1pt}{.1pt}}
\put(411,492){\rule{.1pt}{.1pt}}
\put(440,489){\rule{.1pt}{.1pt}}
\put(469,486){\rule{.1pt}{.1pt}}
\put(498,484){\rule{.1pt}{.1pt}}
\put(528,482){\rule{.1pt}{.1pt}}
\put(557,481){\rule{.1pt}{.1pt}}
\put(586,480){\rule{.1pt}{.1pt}}
\put(616,479){\rule{.1pt}{.1pt}}
\put(645,479){\rule{.1pt}{.1pt}}
\put(674,478){\rule{.1pt}{.1pt}}
\put(704,478){\rule{.1pt}{.1pt}}
\put(733,478){\rule{.1pt}{.1pt}}
\put(762,479){\rule{.1pt}{.1pt}}
\put(791,479){\rule{.1pt}{.1pt}}
\put(821,479){\rule{.1pt}{.1pt}}
\put(850,480){\rule{.1pt}{.1pt}}
\put(879,481){\rule{.1pt}{.1pt}}
\put(909,482){\rule{.1pt}{.1pt}}
\put(938,482){\rule{.1pt}{.1pt}}
\put(967,483){\rule{.1pt}{.1pt}}
\put(997,484){\rule{.1pt}{.1pt}}
\put(1026,485){\rule{.1pt}{.1pt}}
\put(1055,487){\rule{.1pt}{.1pt}}
\put(1084,488){\rule{.1pt}{.1pt}}
\put(1114,489){\rule{.1pt}{.1pt}}
\put(1143,490){\rule{.1pt}{.1pt}}
\put(1172,492){\rule{.1pt}{.1pt}}
\put(1202,493){\rule{.1pt}{.1pt}}
\put(1231,494){\rule{.1pt}{.1pt}}
\put(1260,496){\rule{.1pt}{.1pt}}
\put(1290,497){\rule{.1pt}{.1pt}}
\put(1319,499){\rule{.1pt}{.1pt}}
\put(1348,500){\rule{.1pt}{.1pt}}
\put(1377,501){\rule{.1pt}{.1pt}}
\put(1407,503){\rule{.1pt}{.1pt}}
\put(1436,504){\rule{.1pt}{.1pt}}
\sbox{\plotpoint}{\rule[-0.250pt]{0.500pt}{0.500pt}}%
\put(1306,582){\makebox(0,0)[r]{$\Lambda^{(3)}_{\overline{MS}}=300MeV$}}
\put(1350,582){\rule{.1pt}{.1pt}}
\put(264,353){\rule{.1pt}{.1pt}}
\put(293,344){\rule{.1pt}{.1pt}}
\put(323,336){\rule{.1pt}{.1pt}}
\put(352,330){\rule{.1pt}{.1pt}}
\put(381,324){\rule{.1pt}{.1pt}}
\put(411,320){\rule{.1pt}{.1pt}}
\put(440,316){\rule{.1pt}{.1pt}}
\put(469,314){\rule{.1pt}{.1pt}}
\put(498,311){\rule{.1pt}{.1pt}}
\put(528,309){\rule{.1pt}{.1pt}}
\put(557,308){\rule{.1pt}{.1pt}}
\put(586,307){\rule{.1pt}{.1pt}}
\put(616,306){\rule{.1pt}{.1pt}}
\put(645,305){\rule{.1pt}{.1pt}}
\put(674,305){\rule{.1pt}{.1pt}}
\put(704,305){\rule{.1pt}{.1pt}}
\put(733,305){\rule{.1pt}{.1pt}}
\put(762,305){\rule{.1pt}{.1pt}}
\put(791,306){\rule{.1pt}{.1pt}}
\put(821,306){\rule{.1pt}{.1pt}}
\put(850,307){\rule{.1pt}{.1pt}}
\put(879,308){\rule{.1pt}{.1pt}}
\put(909,308){\rule{.1pt}{.1pt}}
\put(938,309){\rule{.1pt}{.1pt}}
\put(967,310){\rule{.1pt}{.1pt}}
\put(997,311){\rule{.1pt}{.1pt}}
\put(1026,312){\rule{.1pt}{.1pt}}
\put(1055,313){\rule{.1pt}{.1pt}}
\put(1084,315){\rule{.1pt}{.1pt}}
\put(1114,316){\rule{.1pt}{.1pt}}
\put(1143,317){\rule{.1pt}{.1pt}}
\put(1172,318){\rule{.1pt}{.1pt}}
\put(1202,320){\rule{.1pt}{.1pt}}
\put(1231,321){\rule{.1pt}{.1pt}}
\put(1260,323){\rule{.1pt}{.1pt}}
\put(1290,324){\rule{.1pt}{.1pt}}
\put(1319,325){\rule{.1pt}{.1pt}}
\put(1348,327){\rule{.1pt}{.1pt}}
\put(1377,328){\rule{.1pt}{.1pt}}
\put(1407,330){\rule{.1pt}{.1pt}}
\put(1436,331){\rule{.1pt}{.1pt}}
\end{picture}
\caption{Renormalization scale dependence of $H_1$} 
\end{figure}
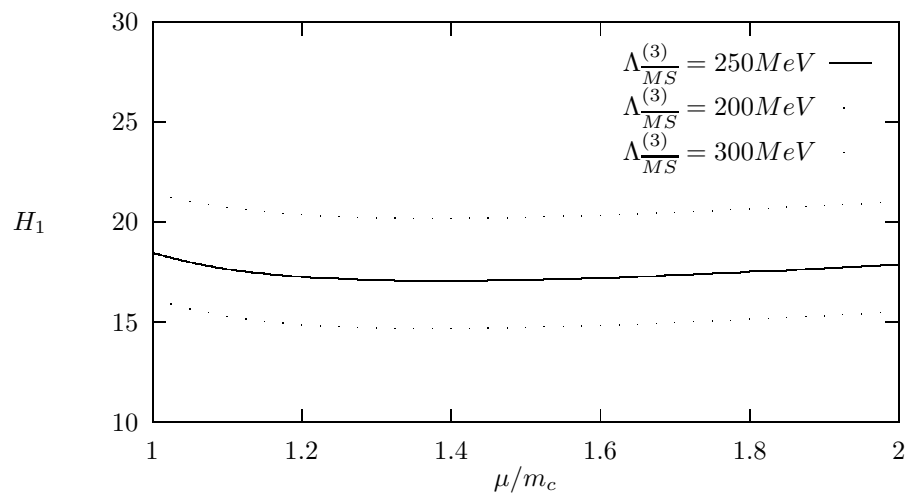

\begin{figure}
\input fig4.tex
\caption{Renormalization scale dependence of $H_8$}
\end{figure}
  
Some comments on the relativistic corrections might be in order. It is obviously
difficult to perform a complete analysis for the $O(v^2)$ corrections, because
it must involve more higher dimensional four-fermion operators whose matrix
elements are difficult to estimate at present. However, we might have some 
phenomenological analyses for the relativistic corrections in the color-singlet
part. Just as the ratio $\Gamma(\chi_{c0}\rightarrow 2\gamma)/\Gamma(\chi_{c2}
\rightarrow 2\gamma)$, which was discussed in ref.\cite{hqc}, the color-singlet 
contribution to $\Gamma(\chi_{c0}\rightarrow 2g)/\Gamma(\chi_{c2}\rightarrow 2g)$
will receive relativistic corrections from two sources, i.e. the kinematic part
and the dynamical part. In the language of the potential model, the color singlet 
matrix element $H_1$ is proportional to $R^{\prime}_P(0)$, the derivative of the
wave function at the origin for the P-states. Due to a strong attractive spin-orbital
force induced by one gluon exchange between quarks, which is also verified by the 
lattice calculations for the spin-dependent potentials between a heavy quark and
an antiquark \cite{michael}, the $\chi_{c0}$ wave function in coordinate space will 
becomes narrower than the $\chi_{c2}$ wave function in which the spin-orbital force
is repulsive, and therefore the derivative of the wave function at the origin becomes
larger for $\chi_{c0}$ than that for $\chi_{c2}$. As a result, the dynamical
relativistic effect will enhances $H_1$ for $\chi_{c0}$ relative to $H_1$ for $\chi_{c2}$,
and this effect is found to be dominant over the kinamatic relativistic corrections
\cite{hqc}. This result might indicate that $O(v^2)$ corrections may further make 
$\Gamma(\chi_{c0}\rightarrow LH)$ enhanced. As for the relativistic corrections in
the color-octet part, more considerations are apparently needed in the future work.

In this paper we give the decay rates of four P-wave quarkonium states into
light hadrons to leading order in $v^2$ and next-to-leading order in $\alpha_s$.
They are expressed in terms of two nonperturbative parameters $H_1$ and $H_8$.
Calculations in dimensional regularization scheme show that the infrared 
divergences, which appeared in the inclusive decay amplitudes for 
$Q\bar{Q}\rightarrow q_l\bar{q_l}g$ and $Q\bar{Q}\rightarrow ggg$, can be 
cancelled explicitly by the contributions of olor-octet operators in NRQCD. 
The finite coefficients of $H_1$ and $H_8$ are given to next to leading order 
in $\alpha_s$. Using the derived theoretical results and the measured decay 
widths of $\chi_{c1}$ and $\chi_{c2}$ we estimate $H_1$, $H_8$ and the decay 
widths of $\chi_{c0}$ and $h_c$.  The determined values are very different from 
the previous values obtained by neglecting the next-to-leading order QCD 
corrections \cite{pwave}. In our results $H_1$ is much larger than $H_8$, and 
the decay width of $\chi_{c0}$ gets enhanced greatly due to $O(\alpha_s)$ 
corrections. As a result, the predicted $\chi_{c0}$ hadronic decay width and 
electromagnetic decay width both could be in agreement with or close to the data. These 
significant differences indicate that QCD radiative corrections are very 
important in understanding the decays of P-wave quarkonium. However, our 
results are valid only to leading order in $v^2$ and next-to-leading order in $\alpha_s$.
The large one-loop coefficients appearing in the expressions of the decay rates indicate that
higher order QCD corrections may be important and that  
our results strongly depend on the choice of renormalization scale. More precise analyses must 
involve relativistic corrections and higher-order QCD corrections, which will include more 
matrix elements of higher dimensional operators.

We would like to thank Professor E. Braaten for pointing out a numerical
error in (\ref{cp2}) and (\ref{3p2}) by comparing their recent result base 
on the threshold expansion method with our result by using the covariant 
projection method in dimensional regularization. It is turned out that the 
two methods in dimensional regularization give identical results for the 
color-singlet sector of the P-wave decay widths, and are consistent with the 
previous calculation of Barbieri $etal.$ using the binding energy as the 
infrared cutoff.

\vfill\eject


\begin{thebibliography}{99}
\bibitem{hag}{R. Barbieri, G. Curci, E. d'Emilio, and E. Remiddi, 
{\it Nucl. Phys.} {\bf B154},535(1979); K.Hagiwara, C.B.Kim, and T.Yoshino, 
{\it Nucl. Phys.} {\bf B177},461(1981).}
\bibitem{Barbi}{R. Barbieri, R. Gatto, and E. Remiddi,
               {\it Phys. Lett.} {\bf 61B}, 465(1976); R. Barbieri, M. Caffo,
               and E. Remiddi, {\it Nucl. Phys.} {\bf B162}, 220(1980); 
              R.Barbieri $etal$., {\it Nucl.Phys.}{\bf B192},61(1981).}
\bibitem{BBL}{G. T. Bodwin, E. Braaten, G. P. Lepage,
		{\it Phys. Rev.} {\bf D51}, 1125 (1995).}
\bibitem{pwave}{G. T.Bodwin, E.Braaten, and G.P.Lepage,
       {\it Phys. Rev.} {\bf 46D}, 1914 (1992).}
\bibitem{huang}{H.W.Huang, K.T.Chao, hep-ph/9601283, to appear in {\it Phys.
 Rev.} {\bf D}.}
\bibitem{mang}{M.L.Mangano and A.Petrelli, {\it Phys. Lett.}
{\bf 352B}, 445(1995)}
\bibitem{3s1}{A.Pertrelli, CERN-TH/96-84 (hep-ph/9603439)}
\bibitem{particle}{Particle Data Group, L. Montanet $et~al$., {\it Phys.
Rev.}{\bf D50(3-I)},1171(1994).}
\bibitem{kuang}{Y.P.Kuang, S.F.Tuan, and T.M.Yan, {\it Phys. Rev.}
{\bf D37}, (1988)1210.}
\bibitem{1p1}{E760 Collaboration (T.A.Armstrong {\bf et~al.}), 
{\it Phys.Rev.Lett.}{\bf 69}(1992)2337.}
\bibitem{E760}{E760 Collaboration (T.A.Armstrong {\bf et~al.}), 
{\it Phys.Rev.Lett.}{\bf 68}(1992)1468;{\bf 70}(1993)2988.}
\bibitem{CLEO}{CLEO Collaboration, J. Dominick {\bf et~al.}, {\it Phys. Rev}.
{\bf D50}(1994)4265.}
\bibitem{chao}{K.T.Chao, Y.B.Ding, and D.H.Qin, {\it Phys.Lett.}
{\bf B301},(1993)282.}
\bibitem{gu}{BES Collaboration, Y.F.Gu {\bf et~al.}, presented at the
Workshop on Beijing $\tau -Charm~Factory$, Beijing, China, Feb.1996.}
\bibitem{hqc}{H.W.Huang, C.F.Qiao, and K.T.Chao, {\it Phys. Rev} {\bf D54},
2123(1996)}
\bibitem{michael}{A.Huntley, C.Michael, {\it Nucl. Phys.} {\bf B286}, 211(1987); 
C.Michael, P.E.L.Rakow {\it Nucl. Phys.} {\bf B256}, 640(1985)}
\end{thebibliography}
\end{document}